\documentclass[onecolumn]{art}
\usepackage{graphicx}
\usepackage{txfonts}

\begin{document}
   \title{Temperature, precipitation and extreme events
during the last century in Italy}

   \author{M. Brunetti
          \inst{1}
          \and
          L. Buffoni\inst{2}
	  \and
	  F. Mangianti\inst{3}
          \and
	  M. Maugeri\inst{4}
	  \and
	  T. Nanni\inst{1}
	  }


   \institute{ISAO-CNR - via Gobetti 101, I-40129 Bologna - Italy\\
         \and
             Osservatorio Astronomico di Brera - via Brera, 28 I-20121 Milan - Italy\\
         \and
	     Ufficio Centrale di Ecologia Agraria - via del Caravita, 7A, I-00186 Roma - Italy\\
         \and
             Istituto di Fisica Generale Applicata - via Brera, 28, I-20121 Milano - Italy\\
             }

 \abstract{
The paper summarises activities within a broad-based research
program for the reconstruction of the evolution of Italian
climate in the twentieth century.
The main result of the program is that Italian climate is
becoming warmer and drier with an increase of both heavy
precipitation events and long dry spells. Most of the
observed signals appear to be due to changes in atmospheric
circulation causing an increase in the frequency of subtropical
anticyclones over the western Mediterranean basin. This
hypothesis is also supported by the evolution of Italian
total cloud amount in the 1951-1996 period.
 }

\maketitle
%

\section{Introduction}

Around the mid 1990s, the authors set up a broad-based research
program with the aim of better understanding the evolution of
Italian climate in the last 100/150 years.
The program was developed both within European (UE IMPROVE
and ALPCLIM projects) and National projects (National Research
Council (CNR) project "Reconstruction of the Past Climate in
the Mediterranean area"). At present it is in progress within
the "Progetto Finalizzato CLIMAGRI", a project of the Italian
"Ministero per le Politiche Agricole e Forestali". Moreover,
in the next two years, further activities will be performed
within the research program "Local climate variability in
relation to global climatic change phenomena" funded by
the Italian "Ministero per l'Istruzione, l'Università e
la Ricerca" and by Genoa, Milan, Trieste, Turin and Udine
Universities. The studies so far carried out have improved
the availability and the quality of Italian data and have
produced interesting information on the evolution of temperature,
precipitation and some other parameters in the last 100/150 years.
The paper summarises the main results obtained within the research
program.


\section{Data and methods}

\begin{figure*}
   \centering
   \includegraphics[width=12cm]{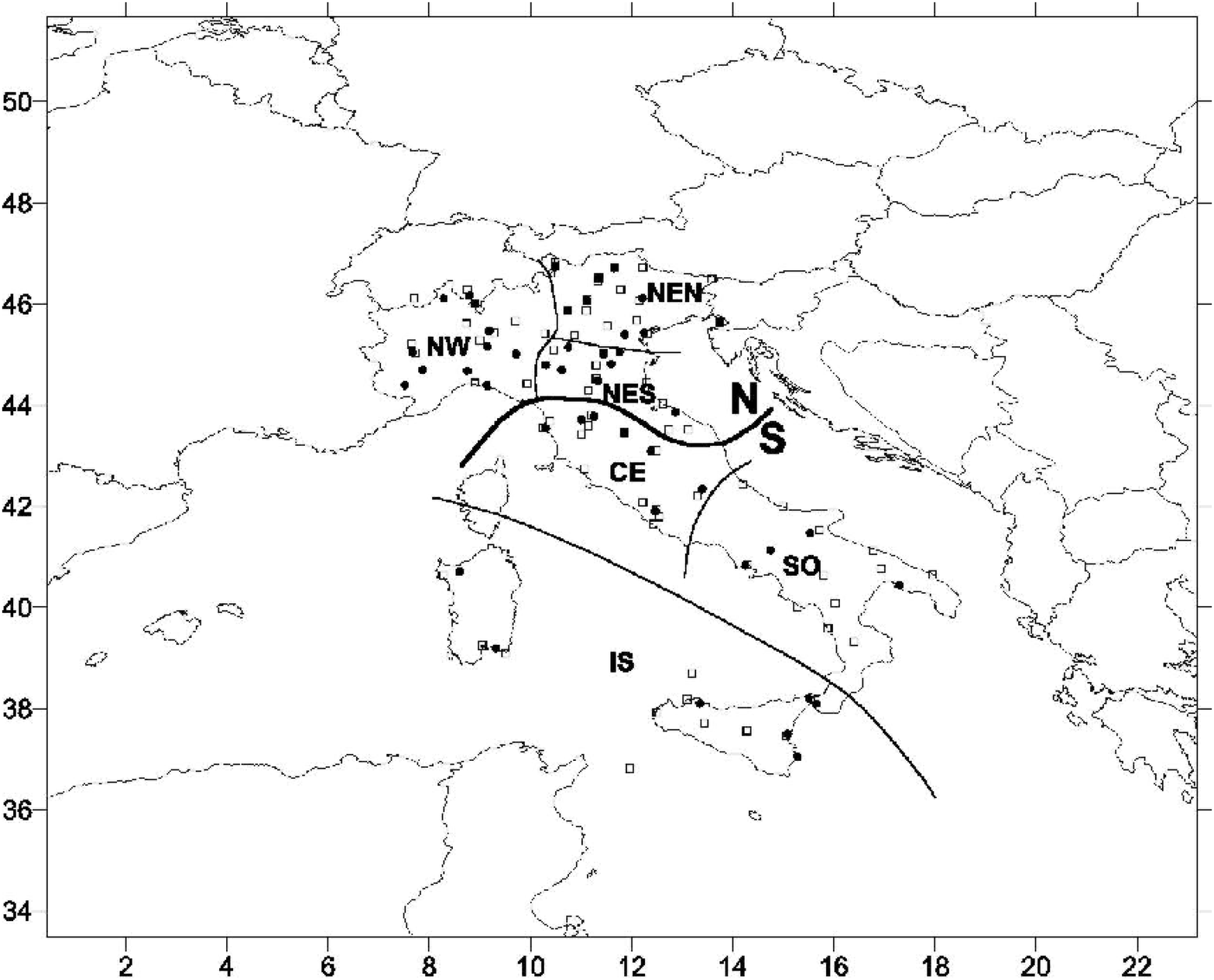}
   \caption{Geographic distribution of the stations.
	Black dots refer to the stations with secular
	series while white squares refer to the stations
	with series available for the last 50 years. Regions
	(N and S) and subregions (NW, NEN, NES, CE, SO, SI)
	are indicated too.}
              \label{mappa}
    \end{figure*}

In Italy there are several historical series, but only a few
of them are digitalized and available for climatological
studies, while most are still on paper archives.
With the exception of some well known series (e.g. Milan,
Padua and Rome), the construction of the most important
database, so far available, of long historical series of
Italian meteorological data started in the 70s in the framework
of a CNR Project. In the second part of the 90s this database
was updated and improved in the framework of the CNR Special
Project "Ricostruzione del clima dell'area mediterranea nel
passato". The series included in the database are listed in
table~\ref{stazioni}, and their geographical distribution is shown in
figure~\ref{mappa}. All series include monthly mean values of daily
maximum (Tmax), mean (T) and minimum (Tmin) temperatures and
monthly total precipitation (P), part of them include daily
observations too.
   \begin{table}
   \centering
   \caption[width=12cm]{List of the historical series,
	co-ordinates of the stations and
	data availability. CNR data base
	is the result of the CNR project
	"Reconstruction of the past climate
	in the Mediterranean area".
	The four last columns of the table
	indicate the goals to be performed
	within 2003 for monthly and daily series.}
   \includegraphics[width=15cm]{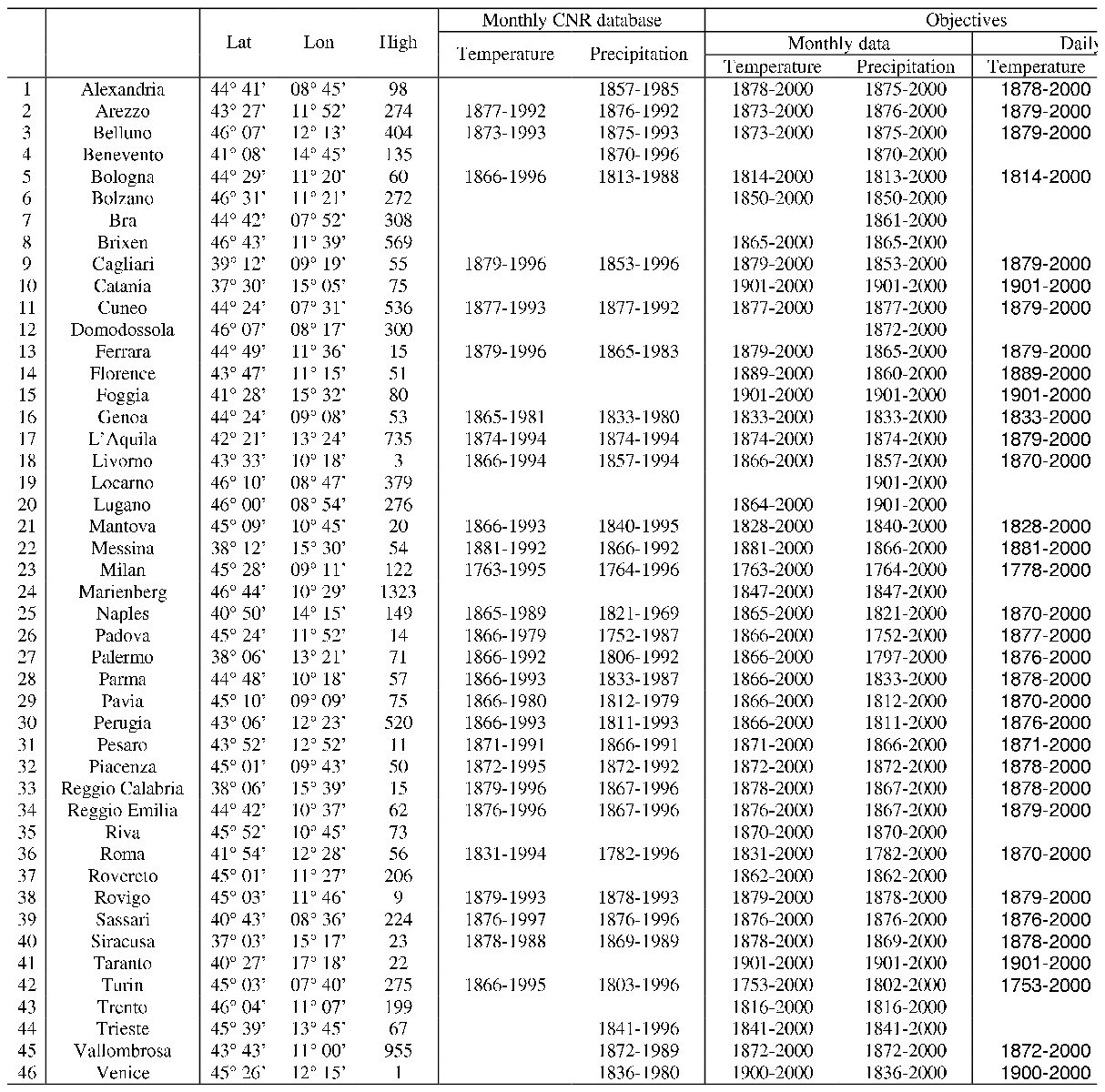}
              \label{stazioni}

    \end{table}

Monthly series were divided in two groups corresponding to
two climatically homogeneous areas - Northern Italy (N) and
Central-Southern Italy (S) - that are, respectively, the
continental and the peninsular zones of Italy
\citep{lovecchio95,buffoni99}.
After establishing the new database, the Craddock homogeneity
test \citep{craddock79} was applied to the Tmin, T, Tmax and P
series, using regional means as reference series
\citep{maugeri98,buffoni99}. Some series were then homogenized
\citep{auer92,boehm92} both on the basis of the test results and
the stations' history (metadata).
After homogenization, the temperature (precipitation) series were
completed over the period 1865-1996 (1833-1996) by means of a
procedure described in \citet{maugeri98} and in \citet{buffoni99}.
With the completed data, monthly mean values of the Daily
Temperature Range (DTR) were calculated from Tmin and Tmax
series. Following the procedure described in
\citet{maugeri98,buffoni99,brunetti00a,brunetti00b} the T, Tmin,
Tmax, DTR and P series were then averaged over N and S and
seasonal and yearly anomalies and their 5-y running means were calculated.
Seasonal and yearly N and S average anomalies were analysed
with the Mann-Kendall non-parametric test, as described in
\citet{sneyers90}, to look for a trend. The slopes of the trends
were calculated by least square linear fitting. The Mann-Kendall
test was also used for a progressive analysis of the series \citet{sneyers90}.
The correlation between seasonal and yearly DTR and seasonal
and yearly precipitation and mean temperature was also performed.
Some secular precipitation series were available with daily
resolution too. For these series the proportion of daily
precipitation falling in 5 precipitation class intervals
was calculated for each year and each season. The resulting
series were then analysed for trends, giving particular emphasis
to heavy and extreme events \citet{brunetti00c}.
The database resulting from the CNR Special Project
"Ricostruzione del clima dell'area mediterranea nel
passato" should be extended to include more series
and further metadata. This extension is currently
underway in the framework of the "Progetto Finalizzato
CLIMAGRI". Other extensions will be performed within
the national research program "Local climate variability
in relation to global climatic change phenomena".
Concerning precipitation, another database, covering a
shorter period (1951-2000), but having a higher spatial
resolution, was considered; most of the series were
extracted from the Italian Air Force (AM) data set,
and some others come from the Ufficio Centrale di
Ecologia Agraria (UCEA), from the Ufficio Idrografico
(UI) and from some research projects dealing with the
recovery of single series. These series were homogenized,
completed, validated and grouped firstly into two areas
(N and S), then in six more restricted areas by principal
component analysis \citet{brunetti01}; the mean
regional series and their anomalies were calculated
for each area. The distribution and the grouping of
these stations are shown in figure~\ref{mappa}. The completion of
the series and the regional means were performed by already
known methods \citep{karl95,karl98,brunetti01}
and by a new method discussed in \citet{brunetti02}.
This new method allows to obtain regional series suitable
for the calculation of statistics (e.g. drought lengths and
frequencies) that can not be obtained using station records
completed by methods based on random generated data.
For the 1951-2000 period, also monthly cloud cover
series, extracted from the Italian Air Force (AM)
data set, were validated, homogenized and analysed for trends.


\section{Results}


\subsection{Monthly series trends}

As far as mean temperature is concerned, on a yearly basis there
is a positive trend with a 0.99 significance level (sl) both for
N and S; on a seasonal basis, considering a 0.99 sl, T has a
positive trend in all four seasons in S, while in N it has a
positive trend in autumn, winter and, using a 0.95 sl, in spring.
The T trends in the annual temperature series, calculated by
least squares linear fitting, range from 0.4 °C/100y for N to
0.7 °C/100y for S. For the winter season the slopes are greater,
ranging from 0.7 °C/100y (N) to 0.9 °C/100y (S), while for the
summer season they are lower and in some cases not significant.
Even if the data set has been checked for homogeneity and some
series have been homogenised, trend estimates seem to be critically
influenced by data homogeneity. The main problem is that data
homogenisation performed on a national basis hardly allows
inhomogeneities due to changes in national standards to be
detected, as they concern most of the series in a rather short
period. The question is still open, as more metadata have to be
recovered before giving a conclusive answer. The information
available at present suggests that the principal inhomogeneities
seem to concern the end of the nineteenth century when in many
observatories the meteorological screens were moved from north
facing windows to more open positions and the end of World War II.
As discussed in \citet{boehm01} the effects of these inhomogeneities
seem to cause an underestimation of temperature trends.
The progressive application of the Mann-Kendall test allows a more
detailed analysis of the temperature series trends. A complete
discussion of this analysis is reported in \citet{maugeri98};
the synthesis of the results is that both for N and S the positive
temperature trend seems to start around 1920. After 1920 the
temperature rises rapidly till 1950, then it is more or less
constant from 1950 to 1985, with only a slight drop in the
period 1970-1980. After 1985 it begins to rise again in all seasons.
Concerning DTR, the results of the analysis are discussed in
detail in \citet{brunetti00b}. The results of the Mann-Kendall
test indicate that DTR has a positive trend (sl $>$ 95\%) with the only
exception of winter in N (negative) and of spring and summer in S
(not significant). The increase in the annual DTR in the period
1865-1996 is weak but significant (0.22 °C for N and 0.12 °C for S)
due to the stronger increase of Tmax compared to Tmin.
The comparison of these results with the literature shows that the
Italian situation is anomalous, because generally the DTR is characterised
by a negative trend \citep{karl93,easterling97}.
A detailed comparison of the results for the period 1865-1996 is
however hampered by the lack of literature data.
A more detailed analysis of the DTR series can be obtained with
the progressive application of the Mann Kendall test. The results
indicate that in the last decades of the 19th century the DTR trend
was generally negative. After the initial decrease, in all the seasons
(except winter) the DTR trend begins to increase from a date included
in the period 1920-1940 for N and in the period 1900-1920 for S.
Then it continues to increase till around 1970 in N and around 1950 in S.
In the last decades the trends are generally constant in N, whereas in S
they are constant in autumn and decreasing in spring and in summer.
In winter there is a tendency towards a negative trend in N and a positive one in S.
As far as precipitation is concerned, the series cover a longer
period than for temperature. However, in order to allow a better
comparison, hereinafter the trend results are presented only for the common period.
On a yearly basis a negative trend (sl 0.99) is evident both for N and S;
on a seasonal basis, there is a negative trend in spring, summer and autumn,
whereas in winter the trend is not significant (S); or it is positive (N).
The slopes of the P yearly series, calculated by least squares linear
fitting, range between -104 mm/100 y for S and -47 mm/100 y for N,
giving estimated decreases in the period 1866-1995 of 135 and 61 mm.
These values correspond, respectively, to 18\% and 7\% of S and N yearly
mean values. Both for N and S, spring and autumn have the steepest trends.

   \begin{table}[htbp]
   \centering
   \caption[]{Yearly and seasonal temperature, precipitation
        and daily temperature range trends for northern and
        southern Italy (period 1867-1996) defined by linear
        regression coefficient (b) and associated error
        (\begin{math} \sigma_b \end{math}).
        Bold numbers: significance level grater than
        99\%, italic numbers: significance level grater than 95\%.}
   \includegraphics[width=8cm]{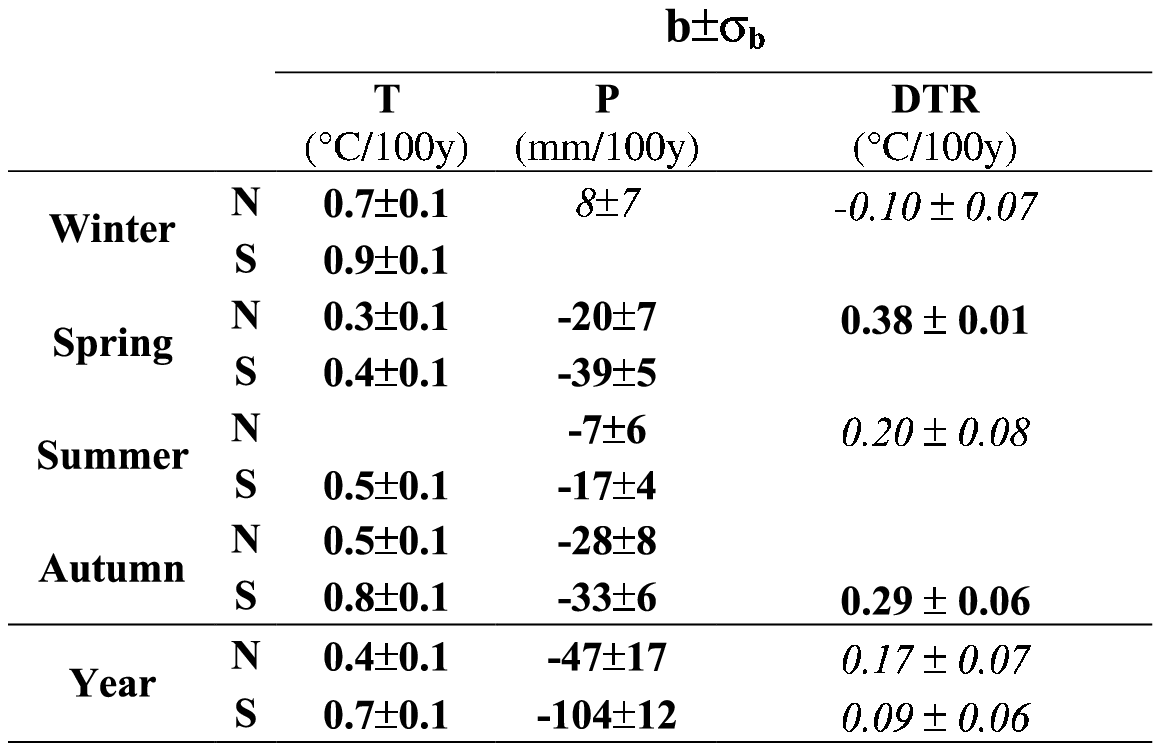}
              \label{trends}

    \end{table}

In the same way as temperature series, also precipitation series were
studied by means of the progressive application of the Mann-Kendall test.
A complete discussion of this analysis is reported in \citet{buffoni99};
the most interesting result regards P in S whose high significant negative
trend seems to be mainly caused by a strong precipitation decrease in the
last 50 years.
All the trend results are summarized in table~\ref{trends}.

\subsection{Relationships among monthly T, P and DTR}

\begin{figure*}[htbp]
   \centering
   \includegraphics[width=15cm]{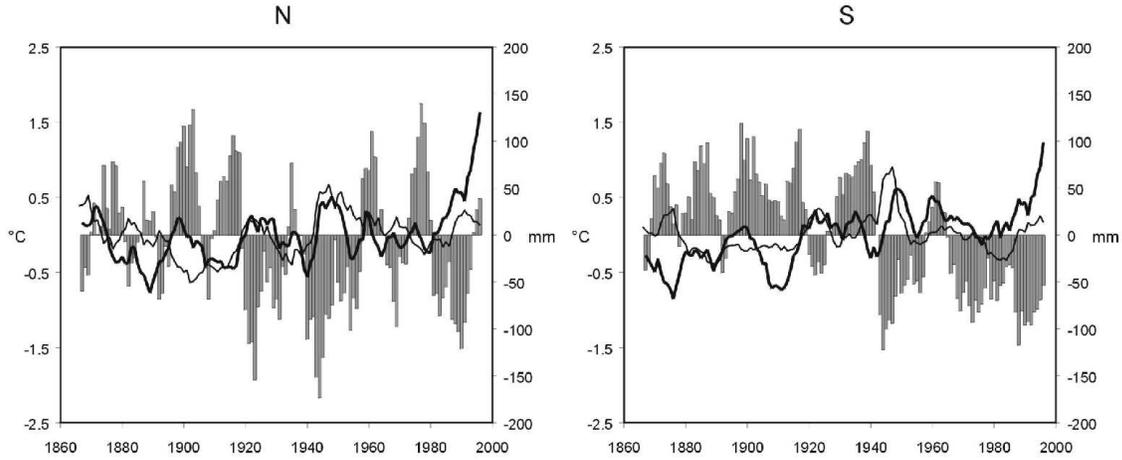}
   \caption{5-year running means of  T (thick line),
	P (histogram) and DTR (thin line).}
              \label{fig2}
    \end{figure*}

The comparison of the behaviour of T, P and DTR for the period 1865-1996
(figure~\ref{fig2}) has been deeply analysed by \citet{brunetti00b}.

The correlation between yearly and seasonal DTR and P is always
negative and highly significant ($>$ 99\%) whereas the correlation
between DTR and T is positive (significance $>$ 95\%) in spring and
summer for N and in spring, summer and autumn for S. The negative
P - DTR correlation is more significant in N than in S whereas the
positive T - DTR correlation is comparable in the two geographical areas.
The correlation among T, P and DTR is mainly due to high frequent variability,
but the same behaviour that is present for the seasonal and yearly data is
evident also on longer time scales. Both the yearly and the secular
correlated behaviours are probably caused by the same changes in
atmospheric circulation, with warm and dry conditions (high T and DTR, low P)
being related to an increase of the frequency of subtropical anticyclones
over the western Mediterranean basin.
These results are also supported by the evolution of Italian total
cloud amount in the 1951-1996 period, which is in very good agreement
with the evolution of the other parameters. By analysing the data of
35 stations we observed that there is a highly significant negative
trend in yearly and seasonal average cloud amount all over Italy
\citep{maugeri01} (table~\ref{wd_pi}). The slope is highest in winter.
   \begin{table}[htbp]
   \centering
   \caption{Results of the application of the Mann-Kendall
	test and of least square linear fitting to the
	regional and subregional WD and PI series.
	In order to allow comparison between the different
	data, the results are expressed as ratios
	(in percentages) between the linear regression
	coefficient and the mean values in the 1951-1996
	period. Bold numbers: significance level greater
	than 95\%; non-bold numbers: significance level
	greater than 90\%; when the significance level is
	lower than 90\% only the sign of the slope is given.}
   \includegraphics[width=15cm]{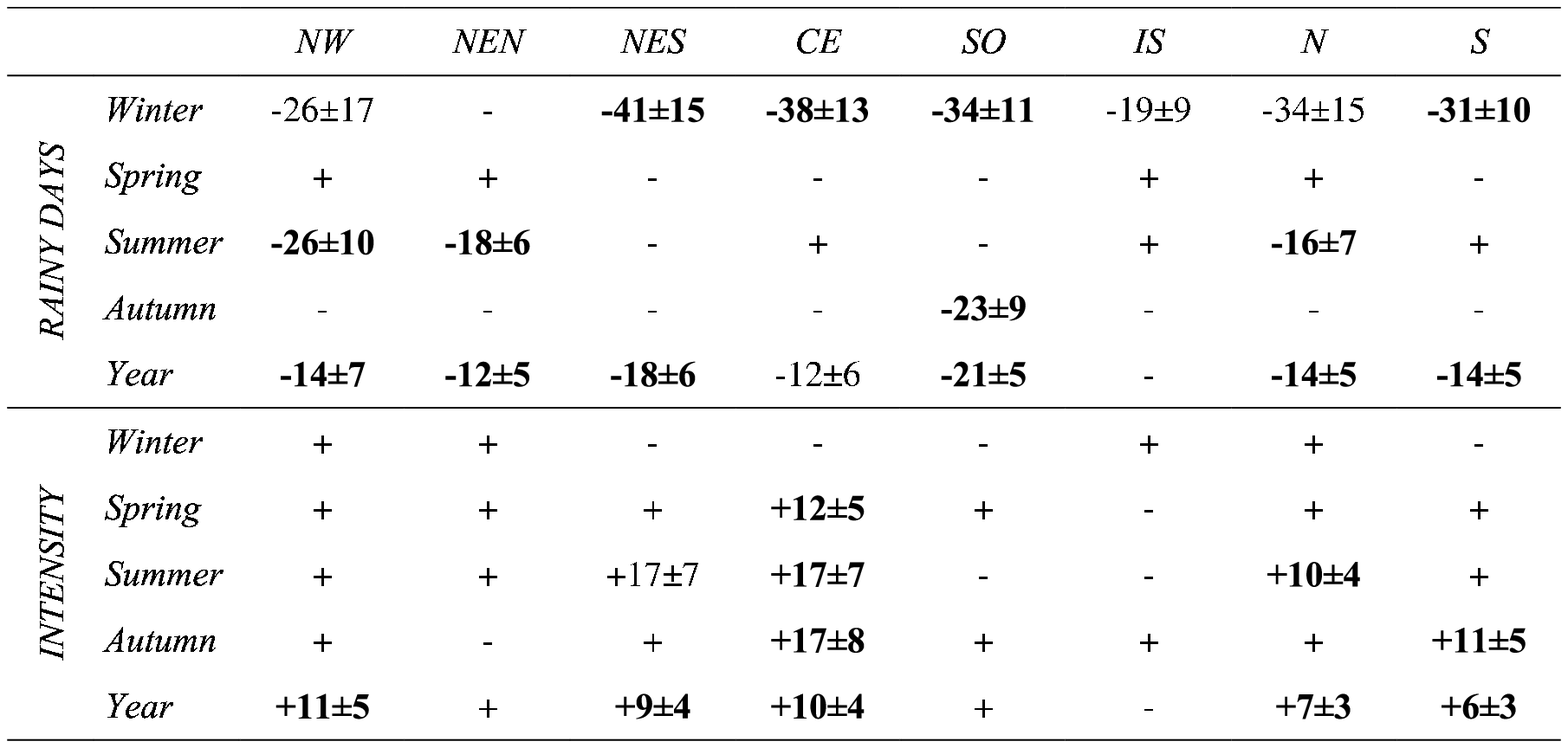}
              \label{wd_pi}

    \end{table}
%


\subsection{Some results concerning daily precipitation series}

An analysis of 67 sites over 46 years of daily precipitation records
for Italy was undertaken by \citet{brunetti01} to identify any
changes in the characteristics of precipitation that may have occurred.
The research concerned seasonal and yearly precipitation, number of wet
days (WDs) and precipitation intensity (PI), and consisted of studying
the trends both for the station records and for some different area
average series. PI was analysed both as amount of precipitation per
wet day and by attributing precipitation to 10 class-intervals
\citep{osborn00} removing the influence of variations in the
number of WDs to yield changes in the underlying shape of the WD
amount distribution.

   \begin{figure*}[htbp]
   \centering
   \includegraphics[width=12cm]{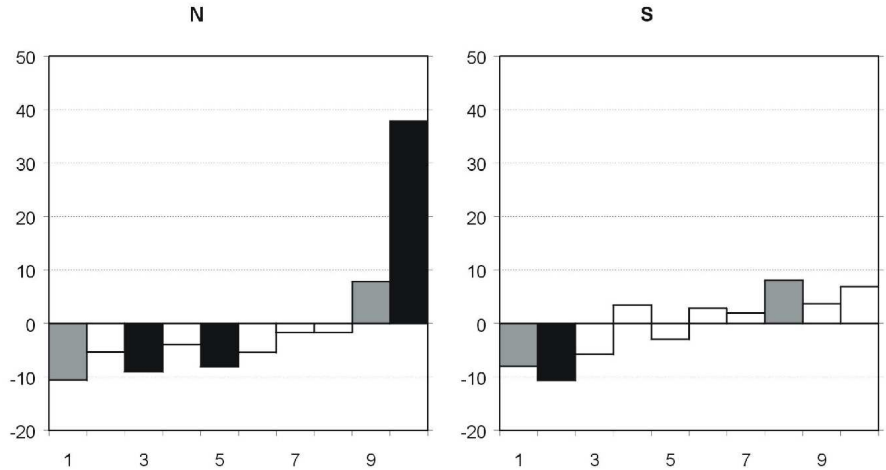}
   \caption{Results of the application of the
	Mann-Kendall test and of least square
	linear fitting to the regional (N and S)
	10 class-interval annual contributions.
	The results are expressed as percentage
	variations compared to the mean value over
	the 1951-1996 period. Black bins: significance
	level greater than 95\%; grey bins:
	significance level greater than 90\%.}
              \label{fig3}
    \end{figure*}
   \begin{table}[htbp]
   \centering
   \caption{Annual and seasonal northern and southern
	Italy cloud amount trends. The results are
	expressed in oktas in 50 years.}
   \includegraphics[width=8cm]{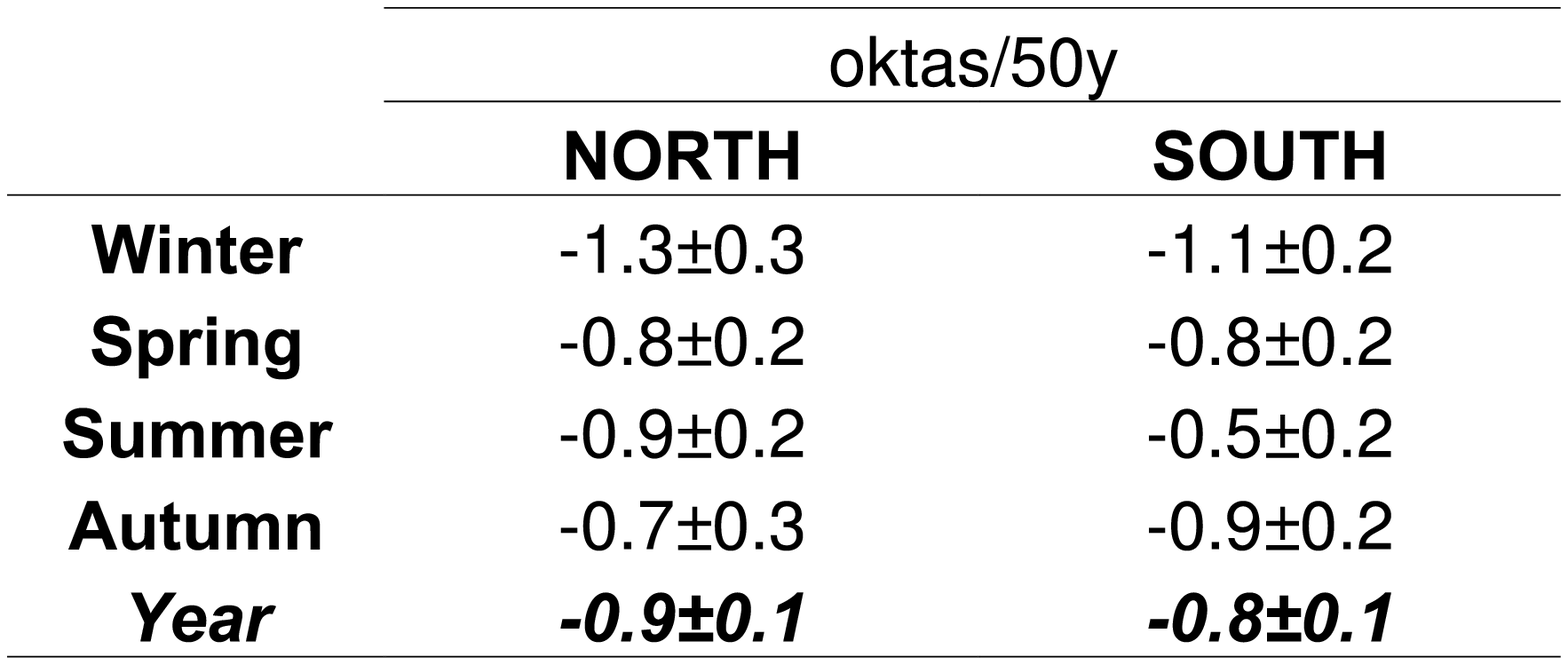}
              \label{cloud}

    \end{table}
The principal results, summarized in figure~\ref{fig3} and table~\ref{cloud}, are as follows:
\begin{description}
\item[i.] The number of WDs in the year has a clear and highly significant
negative trend all over Italy. It depends mainly on the winter, the
season which accounts for about 50\% of the N decrease, and about
75\% of the S one.
\item[ii.] Besides the reduction in the number of WDs, there is a
tendency towards an increase in PI. This increase is globally less
strong and significant than the decrease in the number of WDs, and
it is not concentrated in one specific season. It is worth noticing
that, in winter, the PI trend is positive, but very weak, in N and negative in S.
\item[iii.] In N, the increase in PI is mainly due to a strong increase
in precipitation, falling into the highest class-interval, whereas in S,
it depends on a larger part of the WD amount distributions. As far as the
lower part of the distribution is considered, the situation is more similar,
and both for N and S, the negative trend is best shown for the sum of the
three lower classes.
\end{description}

\begin{figure*}[htbp]
   \centering 
   \includegraphics[width=15cm]{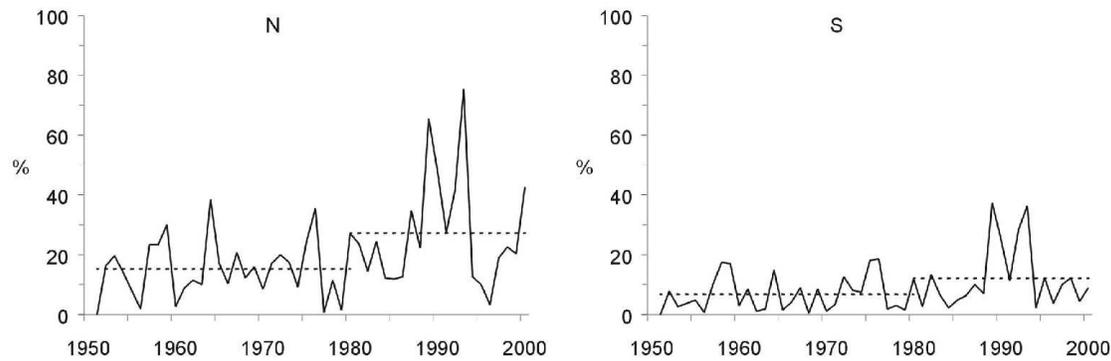}
   \caption{Proportion of winter dry days in N and S.
	In order to highlight the increase in the
	last 20 years, the averages over 1951-1980
	and 1981-2000 (dashed lines) are shown too.}
              \label{fig4}
    \end{figure*}

To search for a trend of dry periods in Italy for the last 50 years,
\citet{brunetti02} proposed a new methods to manage series with
missing data without using random generated numbers and analysed the
previous data set enlarged to 75 stations and updated to 2000. The
stations were clustered as already described for the previous database
and complete daily regional average series were obtained from the incomplete
station records, and analysed for droughts. Droughts were identified by means
of 2 indicators: the longest dry period and the proportion of dry days.
The most remarkable result is a systematic increase in winter droughts
over all of Italy, especially in the North, mainly due to the very dry
1987-1993 period (figure~\ref{fig4}).

The reliability of the regional series was checked by computing some
basic statistics concerning total precipitation, rainy days and
precipitation intensity and comparing them with the same statistics
computed for regional series obtained by station records completed
with methods based on random number generators.


\section{Conclusions}

Globally, our results support the hypothesis we advanced in some
previous papers \citep{brunetti00a,brunetti00b} and confirm that there
are strong differences in the response to the recent global warming
in Italy and other Central and Northern European areas. In fact,
besides some common features (temperature increase, tendency toward
an increase in precipitation intensity), there are other data
(cloud cover, daily temperature range, number of wet days) that
behave conversely. All these differences are probably due to the
strengthening in the North Atlantic Oscillation \citep{hurrell95,hurrell96}
that has accompanied the recent warming and that has caused an
increase and a northerly shift in the westerlies, with consequent
advection of warm and moist air over large areas of Central and Northern
Europe \citep{hurrell95,jones97}, and more frequent anticyclones 
over its southern part \citep{rodwell99}.

\begin{acknowledgements}
The authors would like to thank all the institutions that contributed to the data set.
They are: Italian Air Force, UCEA, Servizio Idrografico, Swiss Meteorological
Institute and CNR-IATA.
\end{acknowledgements}


\begin{thebibliography}{}

  \bibitem[Auer, 1992]{auer92} Auer, I. 1992,
	Experience with the completion and homogenisation of long term
	precipitation series in Austria, Central European Research
	Initiative - Project group Meteorology - working paper 1,
	ed.\ I. Auer (Vienna).

  \bibitem[B\"{o}hm, 1992]{boehm92} B\"{o}hm, R. 1992, Description of the procedure of
	homogenizing temperature time series in Austria, Central
	European Research Initiative - Project group Meteorology - working
	paper 2, ed.\ R. B\"{o}hm, (Vienna).

  \bibitem[B\"{o}hm et al., 2001]{boehm01} B\"{o}hm, R., Auer, I., Brunetti, M., Maugeri, M., Nanni,
	T., Schöner W. 2001, Regional Temperature Variability in the European
	Alps 1760-1998 from homogenised instrumental time series. Int.
	J. Climatol., 21, 1779-1801.

  \bibitem[Brunetti et al., 2000a]{brunetti00a} Brunetti, M., Maugeri, M., Nanni, T. 2000a.
	Variations of temperature and precipitation in Italy from 1866 to
	1995. Theor. Appl. Climatol., 65, 165-174.

  \bibitem[Brunetti et al., 2000b]{brunetti00b} Brunetti, M., Buffoni, L., Maugeri, M., Nanni,
	T. 2000b. Trends of minimum and maximum daily temperatures in Italy
	from 1865 to 1996. Theor. Appl. Climatol., 66, 49-60.

  \bibitem[Brunetti et al., 2000c]{brunetti00c} Brunetti, M., Buffoni, L., Maugeri, M., Nanni,
	T. 2000c. Precipitation intensity trends in Northern Italy. Int. J.
	Climatol., 20, 1017-1031.

  \bibitem[Brunetti et al., 2001]{brunetti01} Brunetti, M., Colacino, M., Maugeri, M., Nanni,
	T. 2001. Trends in the daily intensity of precipitation in Italy
	from 1951 to 1996: Int. J. Climatol., 21, 299-316.

  \bibitem[Brunetti et al., 2002]{brunetti02} Brunetti, M., Maugeri, M., Nanni, T., Navarra
	A. 2002. Droughts and extreme events in regional daily Italian
	precipitation series, Int. J. Climatol., 22, 1455-1471.

  \bibitem[Buffoni et al., 1999]{buffoni99} Buffoni, L., Maugeri, M., Nanni, T. 1999.
	Precipitation in Italy from 1833 to 1996. Theor. Appl. Climatol.,
	63, 33-40.

  \bibitem[Craddock, 1979]{craddock79} Craddock, J. M. 1979. Methods for comparing
	annual rainfall records for climatic purposes. Weather, 34, 332-346.

  \bibitem[Easterling et al., 1997]{easterling97} Easterling, D.R., Horton, B., Jones, P.D.,
	Peterson, T.C., Karl, T.R., Prker, D.E., Salinger, M.J., Razuvayev,
	V., Plummer, N., Jamason, P., Folland, C.K, 1997:   Maximum and
	minimum temperature trends for the globe. Science, 277, 364-367.

  \bibitem[Karl et al., 1993]{karl93} Karl, T.R., Jones, P.D., Knight, R.W., Kukla, G.,
	Plummer, N., Razuvayev, V., Gallo, K.P., Lindseay, J., Charlson
	R.J., Peterson T.C., 1993: Asymmetric trends of daily maximum
	and minimum temperature.  Bull. Am. Meteorol. Soc., 74, 1007-1023.

  \bibitem[Karl et al., 1995]{karl95} Karl, T.R., Knight, R.W. and Plummer, N. 1995.
	Trends in high-frequency climate variability in the twentieth
	century. Nature. 377: 217-220.

  \bibitem[Karl et al., 1998]{karl98} Karl, T.R, Knight, R.W., 1998. Secular trends of
	precipitation amount frequency and intensity in the United States.
	Bull. Am. Met. Soc.. 79: 231-241.

  \bibitem[Hurrell, 1995]{hurrell95} Hurrell, J.W. 1995. Decadal trend in North
	Atlantic oscillation regional temperatures and precipitation.
	Science. 269, 676-679.

  \bibitem[Hurrell, 1996]{hurrell96} Hurrell, J.W., 1996. Influence of variations in
	extratropical wintertime teleconnections on northern hemisphere
	temperature,  Geophys. Res. Lett, 23, 665-668.

  \bibitem[Jones et al., 1997]{jones97} Jones, R.G., Murphy, J.M., Noguer, M., Keen, A.B.
	1997. Simulation of climate change over Europe using a nested
	regional-climate model. II: Comparison of driving and regional
	model responses to a doubling of carbon dioxide, Q. J. R.
	Meteorol. Soc., 122, 265-292.

  \bibitem[Lo Vecchio and Nanni, 1995]{lovecchio95} Lo Vecchio, G., Nanni, T. 1995.
	The variation of the atmospheric temperature in Italy during
	the last one hundred years and its relationship with Solar
	Output. Theor. Appl. Climatol., 51 (3), 159-165.

  \bibitem[Maugeri and Nanni, 1998]{maugeri98} Maugeri, M., Nanni, T. 1998.
	Surface air temperature variations in Italy: recent
	trends and an update to 1993. Theor. Appl. Climatol., 61, 191-196.

  \bibitem[Maugeri et al., 2001]{maugeri01} Maugeri, M., Bagnati, Z., Brunetti, M., Nanni T.
	2001. Trends in Italian cloud amount, 1951-1996. Geophys. Res. Let.,
	28, 4551-4554.

  \bibitem[Osborn et al., 2000]{osborn00} Osborn, T.J., Hulme, M., Jones, P.D. and Basnett,
	T.A. 2000. Observed trends in the daily intensity of United Kingdom
	precipitation. Int. J. Climatol.. 20, 347-364.

  \bibitem[Rodwell et al., 1999]{rodwell99} Rodwell, M.J., Rowell, D.P., Folland, C.K.
	1999. Oceanic forcing of the wintertime north atlantic oscillation
	and European climate, Nature, 398, 320-323.

  \bibitem[Sneyers, 1990]{sneyers90} Sneyers, R. 1990. On the statistical analysis
	of series of observation. WMO, Technical Note N. 143, Geneve, 192 pp.

\end{thebibliography}
\end{document}